\documentclass[prl,twocolumn,showpacs,letterpaper,showpacs,superscriptaddress]{revtex4}
\usepackage{graphicx,amsmath,amssymb,amsfonts,latexsym,color,dcolumn,bm}
\usepackage{verbatim}

\begin{document}
\newcommand{\beq}{\begin{equation}}
\newcommand{\eeq}{  \end{equation}}
\newcommand{\bea}{\begin{eqnarray}}
\newcommand{\eea}{  \end{eqnarray}}
\newcommand{\bit}{\begin{itemize}}
\newcommand{\eit}{  \end{itemize}}
\newcommand{\jmax}{j_{\text{max}}}

\providecommand{\abs}[1]{\left\lvert#1\right\rvert}
\providecommand{\norm}[1]{\lVert #1 \rVert}
\providecommand{\moy}[1]{\langle #1 \rangle}
\providecommand{\bra}[1]{\langle #1 \rvert}
\providecommand{\ket}[1]{\lvert #1 \rangle}
\providecommand{\braket}[2]{\langle #1 \rvert #2 \rangle}

\title{Thermal entanglement in a molecular chain}

\author{P\'erola Milman} \email[]{perola.milman@u-psud.fr}\affiliation{Laboratoire de Photophysique Mol\'{e}culaire du CNRS, Univ. Paris-Sud 11, 
B\^{a}timent 210--Campus d'Orsay, 91405 Orsay Cedex, France}
\author{Arne Keller}\affiliation{Laboratoire de Photophysique Mol\'{e}culaire du CNRS, Univ. Paris-Sud 11, 
B\^{a}timent 210--Campus d'Orsay, 91405 Orsay Cedex, France}

\begin{abstract}
We investigate entanglement  in a linear chain of $N$ polar molecules coupled by dipole interaction.  In our model,  nearest neighbour interaction  predominate, and we compute entanglement  with the help of a two-party correlation entanglement measure. We find that, in this system, only excited states are entangled. Moreover, when  an electrostatic field is applied, energy levels crossings occur,  changing significantly the system's entanglement properties. We make a systematic study of the entanglement dependency on the inter molecular distance separating pairs of molecules, different partitions of the chain and physical parameters as the temperature and the electrostatic field's intensity, showing that it persists for relatively high temperatures and changes its nature with varying field. \end{abstract}

\maketitle

The physics of quantum many-body interacting systems is a passionating research field in the frontier of quantum mechanics, statistical physics and quantum field theory. About $30$ years ago, quantum many-body systems have also been implicated in the domain of quantum information theory. At its origin itself \cite{FEYNMAN}, quantum computers were conceived as physical systems allowing for the simulation of complex quantum many body systems in a controllable context, working as analogue computers. Since then, the interlace between quantum many-body systems and quantum information, has increased and showed itself as a new manner of looking at quantum phase transitions (QPTs)\cite{VEDRAL}, as well as a major application of controllable physical systems, as trapped ultracold atoms and molecules \cite{MOTT, ZOLLER}. Pioneer works studied the behaviour of a particular entanglement measure, the concurrence ${\cal C}$ \cite{WOOTTERS}, close to a QPT in a chain of spins $1/2$ interacting via the Ising Hamiltonian \cite{OSTERLOH}, showing  that  singularities in the derivative of ${\cal C}$ also appear in the thermodynamic limit. Long range spin models  were equally  studied from the entanglement point of view \cite{REMY}, and it was shown that discontinuities also occur for entanglement  measures close to QPT points. An application of such studies concerns collective physical properties of spin systems, as magnetic susceptibility, that can be directly influenced by the presence of entanglement  \cite{NATURE, VEDRAL2}.  In parallel to that, an important step has been taken in the domain of simulation of quantum many-body models with the advent of optically trapped cold atom and molecular physics  \cite{DALIBARD, MOLECULES, ZOLLER}. Theoretical models  show that ultra cold atoms posed in optical lattices can simulate Bose-Hubard spin models \cite{CIRAC}. Experimentally, the observation of the Mott-insulator transition marked the beginnings of the domain of solid state physics with cold atoms \cite{MOTT, DALIBARD}. Technical developments  turned possible the cooling and trapping of more complex structures, as diatomic molecules \cite{MOLECULES2}, that present some potential advantages with respect to atomic systems: polar molecules can have a strong dipole moment, controlled by manipulating the molecular internal states. This can be used as a toolbox to simulate solid state spin models  \cite{ZOLLER}. 

In the present paper, we develop a new way of looking at trapped cold polar molecules, different from the quantum simulation perspective. Since an ensemble of polar molecules is naturally a many-body interacting system, we describe its Hamiltonian's properties using techniques similar to those employed for treating entanglement in solid state spin systems.  We deal here with a high spin-like system: the value of $j$, the single molecule angular momentum, determines the dimension of each ``spin". Our system interacts at its own manner, via dipole-dipole coupling, giving rise to a qualitatively different behaviour than the one observed in condensed matter systems with respect to its singularities. In the present paper, we are mainly interested on the entanglement  in different types of eigenstates of the molecular ensemble and how it can be modified by physical parameters as the intensity of an applied electrostatic field or the temperature $T$.  We also analyse how the variance of the $z$ projection of the total angular momentum $J_z$, $(\Delta J_z)^2$, is affected by the same parameters. This quantity is  analogous to the variance of the magnetisation in spin systems, and can be used to label different types of entangled states, as seen in the following.  Our results have applications in the physics of cold molecules, helping the preparation and the characterisation of different collective states. From the fundamental point of view, it exhibits a number of interesting features, as the presence of thermal entanglement \cite{THERMAL}, long range correlations and a discontinuous behaviour.

 We  deal here with a one dimensional chain of diatomic polar molecules coupled by dipole interaction.  A simplifying assumption made here is that molecules are sufficiently far apart, so that the dipole interaction can be treated as a perturbation coupling nearest neighbouring molecules only. The dipole interaction scales as $1/r^3$, where $r$ is the inter-molecular distance, considered as constant. This means that  interaction strengths lower than  $1/8$th of the one coupling nearest neighbours  are disregarded. As an example, we can take the  KRb molecule, with $\mu=1.2$D and  $B \approx 10$GHz. When, in a chain, KRb molecules are spaced of $r\approx 5$nm, we have that $V_{dip}/2B \approx 5\%$.

 We now move to the Hamiltonian of the system, which determines all the properties we wish to investigate. Each molecule is treated as a rigid rotor and  the $i$-th molecule rotational eigenstate is $\ket{j_i,m_i}$, where $j_i$ is the total angular momentum quantum number and $m_i$ its projection onto the $z$ axis, taken as the chain axis. The free molecule's  energy is  $E_i=j_i(j_i+1)B$, where $B$ is the rotational constant. The total  energy of a chain of $N$ free molecules in the eigenstate $\ket{j_1,m_1}\otimes\ket{j_2,m_2}\otimes...\otimes\ket{j_{N-1},m_{N-1}}\otimes\ket{j_N,m_N}$ is $E=B \sum_i^N j_i(j_i+1)$. We have here  a highly degenerated system: for each molecule, for a given $j_i$, there are $2j_i+1$ states with different $m_i$. When considering a chain of $N$ molecules,  we add  degeneracies coming from  all the permutations of the different $j_i$'s with the same energy.
 
We now describe the dipole interaction, that may lift part of the degeneracy. By supposing that all the molecules have the same dipole moment $\vec \mu$, the dipole-dipole interaction between two neighbouring molecules is $V_{dip}=|\mu|^2/(2\pi \epsilon_o r^3)(- 2\cos{\theta_i}\cos{\theta_{i+1}}+\sin{\theta_i}\sin{\theta_{i+1}}\cos{(\phi_{i}-\phi_{i+1})})$, where $\theta_i$ is the angle the $i$-th molecule makes with the $z$ axis  and $\phi_i$ is the corresponding azimuthal angle. $V_{dip}$ possess  symmetries and selection rules simplifying the  treatment of the problem.  For a given pair of molecules, the dipole interaction has a non null matrix element $\bra{j'_i,m'_i}\bra{j'_{i+ 1},m'_{i+1}}V_{dip}\ket{j_i,m_i}\ket{j_{i+1},m_{i+1}}$ only if $j'_k=j_i\pm 1$ and $m_i+m_{i+1}=m'_i+m'_{i+1}$.  The interest of $V_{dip}$ is that it creates entanglement. To see how it happens, we  consider the simple case of two molecules, that  helps giving  an insight of what happens for  $N \gg 2$ molecules. The free Hamiltonian ground state is trivial, non-degenerated, and non-interacting.  We  then discuss the first excited state, $6$ times degenerated, of energy $E=2B$, and composed by states $\ket{1,0}\ket{0,0}$, $\ket{1,1}\ket{0,0}$ and $\ket{1,-1}\ket{0,0}$ and all the other three permutations. $V_{dip}$  creates a new eigensystem, composed of states $\ket{\Psi_{\pm}^{0}}=1/\sqrt{2}(\ket{1,0}\ket{0,0}\mp\ket{0,0}\ket{1,0})$, of energy ${\cal E}^{(0)}_{\pm}=2B \pm |\mu|^2/(6\pi \epsilon_o r^3)$ and the degenerated states $\ket{\Psi_{\pm}^{1}}=1/\sqrt{2}(\ket{1,1}\ket{0,0}\pm\ket{0;
,0}\ket{1,1})$ and  $\ket{\Psi_{\pm}^{-1}}=1/\sqrt{2}(\ket{1,-1}\ket{0,0}\pm\ket{0,0}\ket{1,-1})$, of energy ${\cal E}^{(1)}_{\pm}=2B \pm |\mu|^2/(12\pi \epsilon_o r^3)$. The energy difference between both types of states occurs because for  $\ket{\Psi_{\pm}^{0}}$ states, the angular probability distribution is maximum on the $z$ axis, corresponding to $\theta_i=0$, minimising $V_{dip}$. In contrast, for $\ket{\Psi_{\pm}^{\pm 1}}$ states, the maximum lies in a plane orthogonal to the $z$ axis.  The first excited state of the dipole interacting system,  $\ket{\Psi_{-}^{0}}$, is pure and entangled. To measure the amount of entanglement of this state, we use the logarithmic negativity \cite{LOG}, defined as ${\cal L}=\log_2{(2{\cal N}+1)}$, where ${\cal N}$ is the negativity, the sum of the negative eigenvalues of the partial transposed density matrix of the two party system \cite{PPT}. In such effective $4$ level system, ${\cal L} \neq 0$ is a sufficient condition for detecting entanglement.  We find  that ${\cal L}=1$ for  $\ket{\Psi_{-}^{0}}$. We can also calculate $(\Delta J_z)^2$ for this state, finding $(\Delta J_z)^2=0$: the angular momentum is orthogonal to the $z$ direction, as mentioned previously. The next lowest energy states are doubly degenerated: $\ket{\Psi_{-}^{\pm 1}}$, with energy ${\cal E}^{(1)}_{-}$. To study their physical properties we assume that they form an equally weighted mixed state. The resulting state, even if mixed, is also entangled, with ${\cal L}=0.77$. We also have $(\Delta J_z)^2=1$,  essentially different from the previous case.  We can thus distinguish the two lowest energy one-excitation states from the point of view of entanglement, $ (\Delta J_z)^2$ and purity, showing that two classes of states can be defined in the subspace with one rotational excitation. 

We now discuss qualitatively the case of a chain of $N$ molecules before moving to our results. Initially,  the system's degeneracy of the first excited level is  $3N$. Among this states, we distinguish $N$ with one excitation of the type $\ket{1,0}$ (called from now on the ${\cal H}_{0}$ subspace) and $2N$ with one excitation of the type $\ket{1,\pm 1}$ (called from now on the ${\cal H}_{1}$ subspace).  When taking $V_{dip}$ into account, states in the ${\cal H}_{0}$ subspace become completely non-degenerated, forming $N$ different entangled states. In contrast, since $V_{dip}$ is independent on the sign of $m_i$, each energy level in the ${\cal H}_{1}$  subspace is two times degenerated, so that entanglement in this subspace is essentially different than in  ${\cal H}_{0}$. Also,  $(\Delta J_z)^2=1$ always in  ${\cal H}_{1}$  and null in ${\cal H}_{0}$.

From this basis, we study now how the spectrum of the system, as well as its properties, can be modified by the application of an electrostatic, linearly polarised, uniform field  in the $z$ direction, interacting independently with each molecule.  The field's interaction Hamiltonian is $V_e=-\sum_i^N \mu_i E_z\cos{\theta_i}$, where $E_z$ is the field amplitude.   We start by analysing its effects over each molecule independently, ignoring $V_{dip}$.  Due to the field's selection rules, single molecule states satisfying $j'_i=j_i\pm1$ and $m'_i=m_i$ only are coupled. However, instead of considering the exact field's eigenstates, we project them into the subspace that is coupled at first order by $V_{dip}$. In this subspace, the   field's  eigenstates are $\ket{+}_i=\cos{\varphi_i}\ket{1,0}_i+\sin{\varphi_i}\ket{0,0}$, $\ket{-}_i=\cos{\varphi_i}\ket{0,0}_i-\sin{\varphi_i}\ket{1,0}$, with  $\cos{\varphi_i}=\sqrt{(B+\lambda)/(2\lambda)}$, and $\lambda=\sqrt{B^2+E_z^2/3}$. The total energy of  $\ket{+}$ and $\ket{-}$ is  $E_{\pm}=B\pm \lambda$. States of the type $\ket{1,\pm1}$ remain unchanged (after projection) but their energy is shifted. Such approximation has been  corroborated by including the $j_i=2$ states in numerical calculations.

\begin{figure}

\includegraphics[width=0.5\textwidth]{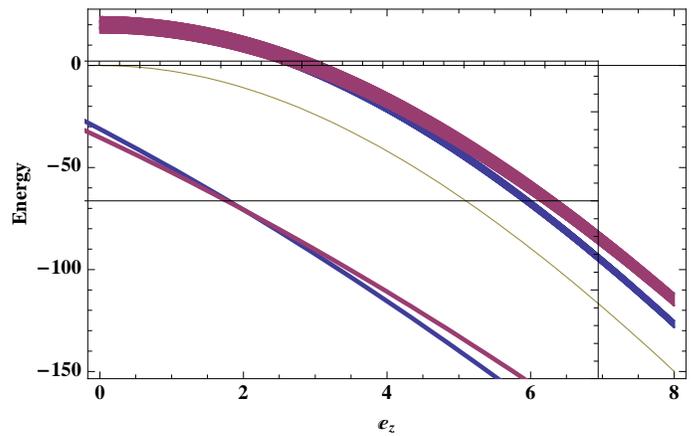}
\caption{(Color online) Energy spectrum of the ground state (brown line) and first excited states for$N=50$ molecules. The ${\cal H}_{+}$ subspace is represented  in violet and the  ${\cal H}_{1}$ one  in blue. All states in the same subspace are equally coupled to the field, and this coupling is different for each subspace, leading to energy levels crossings until the  ${\cal H}_{1}$ subspace becomes the ones with lower energy. In the inset, the crossing between the lowest energy state in subspace   ${\cal H}_{+}$ (violet) and the lowest energy state in the   ${\cal H}_{ 1}$ one (blue).  
\label{fig1}}
\end{figure}

In a second step, we consider a molecular chain, and  compute the effects of  $V_{dip}$.  To do so, the best strategy is to express $V_{dip}$ in the basis of the field eigenstates. In this case, the system's Hamiltonian is tri-diagonal and can be analytically solved \cite{EXACT}. We find then that the ground state is still  non entangled in first order. The first excited level has $N-1$ molecules in state $\ket{-}$ and one molecule in one of the three possible excited states, $\ket{+}$, $\ket{1,\pm1}$. Again, two subspaces can be defined, the one with one $\ket{+}$ excitation (${\cal H}_{+}$) and the one with one $\ket{1,\pm 1}$ excitation  (${\cal H}_{1}$).  The system's spectrum for $N=50$ depends on  the dimensionless electric field amplitude $e_z=E_z 4 \sqrt{3}\pi \epsilon_o r^3/|\mu|$ as shown in Fig. \ref{fig1}. There are no level crossings inside a same subspace,  since states in a same subspace are coupled to the field in the same way. On the other hand, the  ${\cal H}_{1}$ subspace is differently coupled to the field than the   ${\cal H}_{+}$ one,  leading to level crossings between the two subspaces. Because each subspace has different $J_z$ values, experimental characterisation of each subspace can be done by polarisation dependent spectroscopy. 

\begin{figure}

\includegraphics[width=0.5\textwidth]{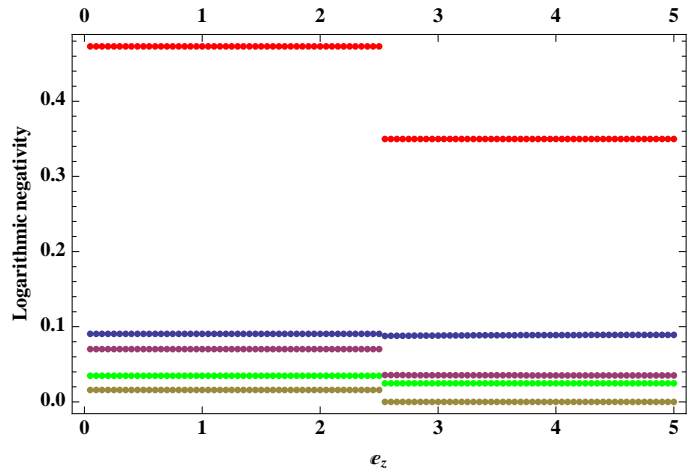}
\caption{(Color online) Logarithmic negativity $L_d$ (blue, pink and brown dots) and $L'_p$ (red and green dots) for different  distances $d$ between a pair of molecules and different positions $p$ of a molecule in a chain, as a function of $e_z$, for the lowest energy state with one rotational excitation. For blue dots, $d=1$, for pink ones, $d=10$ and for brown ones, $d=25$. Red dots show $L_p$ for $p=26$ and green ones for $p=1$. \label{fig3}}
\end{figure}

We study now the observable and relevant  consequences such level crossings have in the system's entanglement. We use here two methods to calculate entanglement. In the first one,  the entanglement of a pair of molecules is evaluated by computing ${\cal L}$  after tracing out the other $N-2$ molecules of the chain.  We can then obtain  $L_d$, the total value of ${\cal L}$, summed over all pairs of molecules at a same relative distance $d$. This is useful for giving  information on the range of entanglement and on how it is distributed over the chain.  A second method to compute entanglement consists of calculating $L'_{p}$, the entanglement of one molecule to the rest of the chain. This quantity, in this  finite  model, depends on the chosen molecule's position $p$: it is maximised for the central molecule and minimised for the extrema ones.We observe that states in the same subspace share the same value of entanglement using such definitions. In  Fig. \ref{fig3} we show $L_d$ as a function of $e_z$ for the lowest energy excited level in a chain of $N=50$ molecules. We see that at the field amplitude corresponding to the energy crossing point, entanglement  suffers a discontinuous change. Also, for each class of state, entanglement does not significantly depends on $e_z$. We observe that entanglement decreases as the distance between pairs of molecules $d$ increases, becoming close to zero at the extrema of the chain.  Discontinuity in $L_d$ is observed for all values of $d$, which is a particularity of this system when compared to other nearest neighbours models \cite{OSTERLOH}. In the same figure, $L'_{26}$ and $L'_{1}$  are plotted for a chain of $50$ molecules, and both quantities suffer a discontinuous change at the energy levels crossing point.

\begin{figure}

\includegraphics[width=0.5\textwidth]{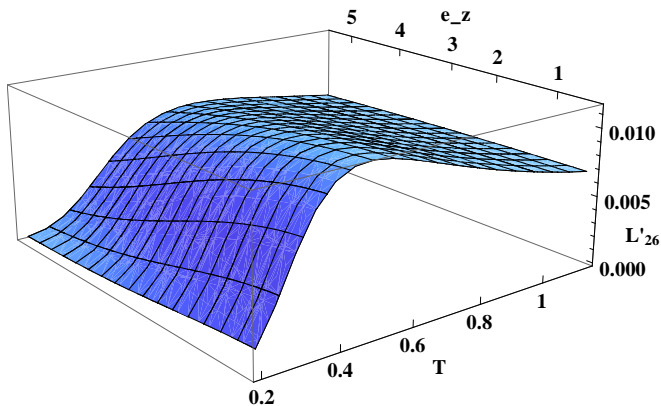}
\caption{(Color online) $L'_{26}$  as a function of ${\bf T}$ and  $e_z$. \label{fig4}}
\end{figure}

In the present system, entanglement is different from zero only for excited states. Consequently, it  displays thermal entanglement: starting from the ground state, entanglement is created by raising the temperature $T$ \cite{THERMAL}. A natural question is how entanglement in thermal equilibrium depends on $e_z$ and on a re-scaled temperature ${\bf T}=k_B T/B$ , where $k_B$ is the Boltzmann's constant. For low enough values of ${\bf T}$, it is reasonable to suppose that only the subspace with one rotational excitation is populated. Fig. \ref{fig4} shows   $L'_{26}$  as a function of ${\bf T}$ and $e_z$ for ${\bf T}$ varying from $0.2$ to $1.2$, which correspond to $T$ varying from $0.1$K to $0.6$K for KRb molecules. We observe that for all ${\bf T}$, the entanglement decreases when $e_z$ increases. This happens because of the level crossings that exchange the energy ordering  of the subspaces ${\cal H}_+$ and ${\cal H}_{1}$. We recall that states in  ${\cal H}_{1}$ are less entangled than the ones in the ${\cal H}_{+}$ subspace (see Figs. \ref{fig1} and \ref{fig3}). We also notice that this entanglement field dependency is less pronounced with increasing ${\bf T}$. This happens because the population difference between each energy level becomes negligible, and we approach a situation where all entangled eigenstantes are equally populated and the system's state is separable. Finally, Fig. \ref{fig4} shows that entanglement reaches a maximum as a function of ${\bf T}$ for all $e_z$. At this point, the excited levels are significantly populated when compared to the ground (non-entangled) state, and at the same time, there is a non-negligible variation of population among them, so that entanglement properties are more pronounced. 
 
In the present paper we have  focused on the   first rotationally excited level, but our results can  be generalised to  highly excited subspaces. In such cases, subspaces with different entanglement properties and differently coupled to the electrostatic filed can always be defined. Consequently, level crossings occur,  modifying  the entanglement and the physical properties of the system.

In conclusion, we have made the first theoretical study of entanglement properties in a chain of polar molecules interacting by dipole force in the presence of an applied static field and with varying temperature. We have identified two subspaces that can be labelled by their $(\Delta J_z)^2$ expectation values.  Our results show  a number of interesting and novel features, as a discontinuity on entanglement for the first excited levels when the electric field's amplitude is varied, irrespectively of the dimension of the chain. We have also investigated how  pairwise entanglement depends on the distance between molecules. We showed that it  decreases with distance, but persists for relatively far apart molecules, even if the interaction has been considered between nearest neighbours only. Moreover, we have calculated entanglement of one molecule at different sites to the rest of the chain, showing that it also presents a discontinuity when $e_z$ is varied.  A striking feature of this system is that entanglement is absent at the ground state but increases with temperature. We have thus studied how this entanglement temperature and field dependency takes place. The present study is relevant for fundamental reasons, since it consists of a many body system exhibiting peculiar entanglement properties, and also for experimental applications. The field of cold molecules trapped in optical lattices has made astonishing progress in the last years, and the results presented here can be useful for the preparation and characterisation of entangled molecular states even  at finite temperature.  The present work opens the perspective of controlling entanglement properties of molecular chains by adjusting external parameters as the temperature and external applied fields. 
 
It is a pleasure to acknowledge Peter Zoller, Andrea Micheli, Tatiana Rappoport and Osman Atabek for fruitful discussions.

\end{document}